# Cloud computing-based higher education platforms during the COVID-19 pandemic

Short Title Cloud computing-based higher education platforms

Hui Han

Data Science Department, Fraunhofer Institute for Experimental Software Engineering IESE, Kaiserslautern, Germany, hui.han@alumnos.upm.es

Silvana Trimi

Department of Supply Chain Management and Analytics, College of Business, University of Nebraska - Lincoln, Lincoln, USA, silvana@unl.edu

Cloud computing has become the infrastructure that supports people's daily activities, business operations, and education delivery around the world. Cloud computing-based education platforms have been widely applied to assist online teaching during the COVID-19 pandemic. This paper examines the impact and importance of cloud computing in remote learning and education. This study conducted multiple-case analyses of 22 online platforms of higher education in Chinese universities during the epidemic. A comparative analysis of the 22 platforms revealed that they applied different cloud computing models and tools based on their unique requirements and needs. The study results provide strategic insights to higher education institutions regarding effective approaches to applying cloud computing-based platforms for remote education, especially during crisis situations.

CCS CONCEPTS • Applied computing • e-Education • e-Learning

**Additional Keywords and Phrases:** Cloud computing, multiple-case study, online education platforms, COVID-19, higher education

## 1 INTRODUCTION

The blunt impact of the COVID-19 pandemic has disrupted every fabric of people's lives from their daily routines, the way they work, learn, socialize, and entertain [5]. One of the long-lasting impact areas of COVID-19 has been education where millions of students had to learn remotely online or through Zoom [7]. China is the country that suffered the initial blow of the pandemic but has successfully managed it [17]. One of the key success factors for China's fast return to almost normal life routines is the aggressive and innovative applications of advanced digital technologies to facilitate online education of 30.3 million Chinese university students in 2019 [13]. Specifically, the Chinese government has successfully applied cloud computing-based platforms to facilitate online teaching and learning for its massive university community. An in-depth analysis of China's online education systems at the university level will provide valuable insights to IT professionals who design service platforms and administrators of higher education worldwide. In this study, we conduct multiple-case analyses to explore how cloud computing-based online higher education platforms were developed and applied in China during the pandemic.

The Chinese Ministry of Education (MOE) announced that the start of the 2020 spring semester of all schools, from kindergartens to universities, across the country would be postponed to control the further spread of the pandemic [2, 18]. Although schools were closed, MOE issued instructions to higher education institutions (HEIs) to deploy online teaching to enable university students to resume their studies remotely. As of February 2020, there were 22 online education platforms providing 24,000 online university-level courses free of charge, covering 12 major fields of studies at the undergraduate level and 18 disciplines at the higher vocational education level (The State Council, 2020; S. Zhang et al., 2020).

In this paper, we examine how Chinese universities deployed cloud computing-based platforms to effectively facilitate online education during the pandemic. More specifically, we apply a multiple-case study to compare the 22 online education platforms from seven main aspects: three cloud computing related variables (cloud service model, cloud deployment model, and cloud computing tools) and four education-related high-level metrics such as support for mobile learning, the number of students accessing a MOOC, the number of courses available, the number of schools collaborating in the program, and course types (the results are presented in Table 1). We also provide detailed information about some unique characteristics when different platforms are involved in cloud computing models (cloud service model and cloud deployment model) as well as cloud computing tools.

The rest of the paper is organized as follows. Section 2 presents the impact of the COVID-19 pandemic on education, as well as the online education systems of Chinese universities and 22 online platforms during the pandemic. In Section 3, we introduce the multiple-case study methodology. Section 4 discusses the results of the multiple-case analyses of the 22 cloud computing-based online higher education platforms. We conclude the study in Section 5 by presenting implications of the study results, limitations, and future research needs.

## 2 BACKGROUND

### 2.1 COVID-19 Pandemic

The World Health Organization (WHO) declare COVID-19 a global pandemic on March 11, 2020. Since then, the pandemic has devasted the world in a matter of several weeks, driving many hospitals to the breaking point with the exploding number of COVID patients, disrupting international transportation, prohibiting most crowd-gathering activities (factories, religious services, schools, sporting events, cultural performances, cruise travels, theaters, and even restaurants), soaring unemployment, and forced lockdown of cities, and the like [5].

The gravity of the pandemic's effects is clearly shown by the number of infected people and deaths (over 80 million and more than 2 million fatalities worldwide) [3] and the global economic decline of about 4.3% in 2020. In the US, the booming economy until early 2020 suddenly turned into a recession with the down spiraling financial markets, a massive number of newly unemployed people in excess of 30 million, and shutting down of schools, workplaces, and public events. The estimated economic damage caused by the pandemic in 2020 alone was over $10.3 trillion [16]. A very fortunate thing about the pandemic, if one has the audacity to even say it, is that the disaster occurred in today's digital era where there are advanced scientific and technological tools available to combat the pandemic (Lee and Trimi, 2021). A good comparison can be made between the Great Influenza epidemic of 1918-1920 and COVID-19. The Great Influenza epidemic was the most devastating pandemic in history, which infected more than 500 million people, about one-third of the world population at that time, and deaths of over 50 million [8]. The Economist (2020) predicted that COVID-19 could inflict more than 200-600 million cases and up to 3.7 million deaths worldwide by 2021 if advanced technologies and medical facilities were not available. In the advent of advanced Industry 4.0 technologies and warp-speed innovation

ideas during the pandemic, corporations, governments, research institutions, healthcare providers, communities, and individuals have implemented numerous measures to manage COVID and its economic and social impacts [11].

**2.2  Online Education in China during the Pandemic**

The Chinese municipal and provincial governments, supported by the central government, have adopted a series of unprecedented response measures, such as restricted travels across cities, strict case detection and contact tracing, quarantine of infected persons, testing kit development, etc. [19].

Although schools were closed, Chinese students were able to attend classes online. The Ministry of Education (MOE) has issued a guideline for universities to organize online classes, with 22 online education platforms offering 24,000 courses to students. In February 2020, MOE also launched the National Online Cloud Classroom (http://ykt.eduyun.cn/), a free online learning platform for students studying at home. For students in remote or rural areas with low Internet accessibility, the platform has incorporated streamed content in the TV series "Taking the Same Class" by CETV-4 (China Education Network Television Channel 4) [2, 18].

For the online education of Chinese universities, cloud computing serves as the backbone infrastructure for all online platforms. Cloud computing is the infrastructure that renders new value to the e-learning environment, as educational services can be distributed in a stable and ubiquitous way[6]. In this study, we perform a multiple-case study of 22 innovative online education platforms that were designed to overcome the interruptions experienced by universities during the pandemic.

**2.3  Basic Information on 22 Online Platforms**

ICourse is a university MOOC (Massive Open Online Course) platform that has more than 30 million registered users and over 140 million learners. In order to reduce the impact of the epidemic on the learning of university students, ICourse platform provides free courses and teaching services for colleges and universities. This platform runs more than 8,000 courses, including 916 national open online courses that support 12 major fields for universities and 18 majors in colleges, as well as 3,915 courses offered by 121 renowned universities.

XuetangX is a MOOC platform launched by Tsinghua University in 2013. It developed the virtual private cloud platform "XueTang Cloud" and the teaching tool "Rain Classroom" for assisting online learning. XueTangX had more than 58.8 million registered online learners in March 2020. To effectively minimize the negative impact of the epidemic, XueTangX has provided 3,000 courses on live streaming services. The services offered include 300 basic courses and 900 professional courses from more than 600 universities.

Wisdom Tree Network is a large-scale service platform for the operation of credit courses. There are nearly 3,000-member schools and more than 20 million students who take credit courses. It helps member schools realize cross-school curriculum sharing and credit recognition, and complete cross-school elective courses. Xueyin Online is a new generation of open learning platforms based on the concept of credit banks. It was created jointly by the National Open University and Chaoxing Group, providing a combined service of MOOCs, Credit Bank, and Education Taobao as an integrated learning environment.

Superstar Erya is a general education brand that Superstar company builds. Unlike professional education, general education pays more attention to the growth of students in improving their personality, humanistic feelings, scientific spirit, and social responsibility. Superstar Erya, as a domestic general education learning platform since 2011, has gradually established itself as a high-quality general education brand in the process of serving more than 1,800 colleges. A total of 15.1 million students have studied and obtained credits on this platform. During the epidemic period, Superstar Erya opened all course resources and teaching aid resources for free.

To adopt new education models in medical higher education, People's Medical Publishing House with 56 Chinese medical associations, formed the Chinese Medical Education MOOC Alliance and developed an online platform named "Human Health MOOC". The alliance includes nearly 200 associations, which oversee almost all domestic medical schools. During the pandemic, the Human Health MOOC has provided free online learning services. There are 5 majors, including clinical medicine, nursing, pharmacy, traditional Chinese medicine, and stomatology, that cover nearly 220 high-quality medical courses on this platform. In addition, Human Health MOOC provides SPOC (Small Private Online Courses) platform services to medical schools to help complete online course teaching management for free.

University Open Online Courses (UOOC) Alliance, as the first high-quality MOOC resource-sharing platform that was developed by local universities, aims to integrate quality teaching resources to support higher education in China. Since its establishment, the alliance has grown steadily. The scale of the affiliated universities has continuously expanded to 153-member universities in 29 provinces and 69 cities across the country. The alliance serves nearly 680,000 students. In the face of the severe pandemic situation, UOOC Alliance has been providing 418 courses to the public free of charge.

"CNMOOC" is the official website of the China High-Level University MOOC Alliance. The alliance is an open cooperative education platform voluntarily established by several high-profile universities in China. The purpose of this alliance is to build a high-level, large-scale online open course platform through exchanges, seminars, consultations and collaborations. During the pandemic period, CNMOOC launched an emergency plan to provide a completely free platform and course services including MOOC courses, online live broadcasts, applet applications, online experiments (EduCoder), teaching service support, learning data support, etc. for all domestic universities.

The "Beijing University Quality Course Research Association" (RongYouXueTang) was established in 2015 under the guidance of the Beijing Municipal Education Commission. The purpose of this association is to reform education, develop high-quality shared courses, promote training models, and improve the quality of training.

Chinese Moocs is a Chinese-based MOOC service platform designed to serve global Chinese people and provides users with free courses from well-known universities such as Peking University. All courses are free and open to the world, and most courses start as self-service courses. Colleges and universities can apply for free SPOC platform services, which provide one-to-one course guidance.

UMOOCs is affiliated with the Foreign Language Teaching and Research Press (FLTRP) Co. Ltd. During the epidemic, the UMOOCs platform has exerted its professional and technical capabilities to provide online foreign language teaching and learning solutions for teachers and students nationwide. The solution consists of three parts: student self-learning solution, teacher online SPOC teaching solution, and teacher self-improvement solution.

The GaoXiaoBang platform has all-around teaching-learning- testing- evaluation- practice functions. This platform has been used by more than 1,500-member colleges, 300 enterprises, and 4.9 million students. Course categories have covered 13 subjects such as military, philosophy, economics, law, education, literature, and history, etc. During the pandemic, online courses, technical platforms and consulting services are freely available to teachers and students. Students can complete a specified course required by their schools or choose courses based on their own interests.

ULearing focuses on the in-depth integration of information technology and education. It provides 500-course resources for free covering 12 categories, as well as supports paper materials, online courses, teaching courseware, after-school exercises and an exam question bank for universities and colleges.

PeopleMooc provides the public with high-quality MOOC courses and supports simultaneous learning on multiple terminals such as PCs and mobile phones. The learning progress is controlled by learners themselves. There are 221 courses on the platform, which are continuously updated. Course contents include Maker Lecture, Current Affairs Education, Sports, Health, Chinese Traditional Culture, Innovation and Entrepreneurship, etc.

"Vocational Education Digital Learning Center" (also known as ICVE) is a vocational education digital teaching resource sharing platform operated by Higher Education Press. During the epidemic, ICVE has helped various vocational schools to create an online "cloud" for developing professional education teaching resources, vocational education cloud SPOC, and MOOC.

During the epidemic, some accounting education experts, together with ZhengBao Distance Education Group, decided to provide free course resources on the online learning platform "Chinese Accounting Net School" to universities and colleges across the country. It provides technical support, storage and analysis of learning data, and experts to answer questions. This platform provides 15 professional core course resources led by accounting experts; 8 live broadcast courses for preliminary and annotation test preparation; 4 innovative elective courses, and a job-search training course for college students.

To ensure the continuation of university education in Zhejiang Province during the epidemic, a green channel (Zhejiang Institutions of Higher Learning Online Open Course Sharing Platform) was opened with MOOC + SPOC platforms, opening courses, student learning services, and technical support.

The Anhui Provincial Education Department developed the Anhui Province online course learning center (e-HuiLearning), which is an online learning platform with the provincial high-quality course resource. Teachers on the platform are encouraged to set up new teaching courses to enrich the platform's curriculum resources. The platform has 868 courses, more than 2,000 teachers and over 490,000 registered users by the end of 2019. It has a total of 695 member colleges, including 434 undergraduate colleges and 261 vocational colleges. In addition, the cumulative visits of the platform have reached more than 123.8 million times.

The Chongqing Universities Online Open Course Platform, based on MOOCs, is a localized platform with national and international high-quality course resources. It aims to build an online open learning platform for colleges and universities in Chongqing. This platform has over 400 national-level courses, more than 56,000 active teachers, and in excess of 1 million students. It provides services such as digital course construction and teaching resource library and a school-level private cloud platform for more than 60 universities in Chongqing.

The experimental space (iLab-X) is the first experimental teaching platform, which directly serves students and social learners in China. To promote open sharing and efficient use, iLab-X created national, provincial, and school-level shared application services. This platform includes 2069 experimental projects, covering 255 majors and 1561 courses. Each virtual simulation experiment teaching project presents a complete promotional video, detailed experiment project description, and seamless docking experiment jump, which provide participants with a simple and convenient experiment channel.

EduCoder is built by several domestic universities and large software companies. Zhiqing Technology Company is one of the software companies, which is responsible for the construction and management of the MOOC online open practice (MOOP) teaching system. EduCoder provides 12,000 teaching resources, 650 online courses, and serves more than 1,000 universities. At present, the EduCoder community has provided more than 400 sets of practical course resource packages for dozens of courses such as programming, data structure, operating system, database, electronic technology, cloud computing, big data, artificial intelligence, and blockchain.

## 3 METHODOLOGY

In this study, we use the multiple-case study methodology to compare the 22 online educational platforms based on cloud computing relevant variables (cloud service models, cloud deployment models, cloud computing tools and mobile learning) and education-related variables (number of students, the number of courses, and the number of member institutions) for supporting remote education during the pandemic in China.

### 3.1 Multiple-case Study

Multiple case studies allow exploring for answers to the research question of "how"? The major benefits of this methodology are to understand the similarities and differences among the cases [10]. The multiple-case study is robust and rigorous as cross-case comparisons allow a deep understanding of the effects of contextual variables. These properties of the methodology make it particularly appropriate for the aim of this study [15].

### 3.2 Variables

*3.2.1 Cloud service models*

There are three main types of services offered by cloud computing: infrastructure as a service (IaaS), platform as a service (PaaS) and software as a service (SaaS) [6]. IaaS provides on-demand physical and virtual computing resources, PaaS supports customers with a rich development environment, and SaaS offers complete application functionality.

*3.2.2 Cloud deployment models*

There are four deployment models with derivative variations that address specific requirements: Public Cloud, Private Cloud, Community Cloud, and Hybrid Cloud [9]. The physical cloud infrastructure of the public cloud is located on cloud provider's premises and it is shared by all cloud customers. The infrastructure of the private cloud is provisioned for exclusive use by a single customer. The infrastructure of the community cloud is for the exclusive use of several customers of the same group. A hybrid cloud is a combination of two or more different cloud infrastructures.

*3.2.3 Cloud computing tools*

Cloud computing tools are mostly used for different collaborative purposes and activities. Based on the nature of their utilization in a blended-specific situation, these tools can be categorized into three types: synchronized tools, learning management system (LMS) tools, and social networking tools [1]. Synchronized tools enable students to carry out various synchronous and asynchronous functions such as editing, comment writing, and peer-review sessions. LMS tools consist of using university systems (such as Moodle and Blackboard) to support learners to document, track, and report on various educational activities. Social networking tools are used for interpersonal communication, sharing, and discussions on study topics.

### 3.2.4 Support for mobile learning

Support for mobile learning means the platform can help overcome the current limitations of mobile learning (m-learning) regarding the processing and storage capabilities of the devices used.

### 3.2.5 Education-related variables

Currently, most of the metrics monitored are related to cloud computing, but other education-related high-level metrics should also be considered, such as the number of students accessing a MOOC, the number of courses available, the number of schools collaborating in the program, and course types.

## 4 RESULTS AND DISCUSSION

The results of the multiple-case study can be summarized as shown in Table 1.

Table 1 Multiple-case study results

| Platform | Cloud Service Model | Cloud Deployment Model | Cloud Computing Tools | Support for Mobile Learning | Students Served | Courses Served | Member Institutions |
|---|---|---|---|---|---|---|---|
| ICourse | SaaS | Private Cloud | All 3 tools | Yes | 30000000 | 8000 | 691 |
| XuetangX | SaaS | Private Cloud | All 3 tools | Yes | 58800000 | 3000 | 600 |
| Wisdom Tree | SaaS | Public Cloud | All 3 tools | Yes | 20000000 | 3842 | 3888 |
| Xueyin | SaaS | Public Cloud | All 3 tools | No | 47000890 | 2130 | 4443 |
| Superstar | SaaS | Public Cloud | All 3 tools | Yes | 15100000 | 480 | 1800 |
| Human Health | SaaS | Community | All 3 tools | Yes | 800000 | 220 | 200 |
| UOOC | SaaS | Community | All 3 tools | Yes | 680000 | 418 | 153 |
| CNMOOC | SaaS | Community | All 3 tools | Yes | 536000 | 313 | 134 |
| RongYouXue | SaaS | Private Cloud | All 3 tools | Yes | 100000 | 200 | 25 |
| Chinese Moocs | SaaS | Public Cloud | LMS tools | No | 386324 | 124 | 3 |
| UMOOCs | SaaS | Community | All 3 tools | No | 1200000 | 151 | 166 |
| GaoXiaoBang | SaaS | Private Cloud | All 3 tools | Yes | 4900000 | 1000 | 1800 |
| ULearing | SaaS | Hybrid Cloud | All 3 tools | Yes | 1000000 | 500 | 600 |
| PeopleMooc | SaaS | Hybrid Cloud | All 3 tools | Yes | 20000000 | 221 | 16 |
| ICVE | SaaS | Hybrid Cloud | All 3 tools | Yes | 7638000 | 1273 | 239 |
| Accounting | SaaS | Public Cloud | All 3 tools | Yes | 100000 | 28 | 3000 |
| ZhengBao | SaaS | Public Cloud | All 3 tools | Yes | 100000 | 44 | 3000 |
| ZIHLOOCSP | SaaS | Hybrid Cloud | All 3 tools | Yes | 1994154 | 3720 | 541 |
| HuiLearning | SaaS | Hybrid Cloud | All 3 tools | Yes | 490000 | 868 | 695 |
| CUOOCP | SaaS | Hybrid Cloud | All 3 tools | No | 1000000 | 400 | 60 |
| iLab-X | SaaS | Public Cloud | LMS and SN | No | 20690000 | 1561 | 239 |
| EduCoder | SaaS | Public Cloud | All 3 tools | Yes | 4000000 | 650 | 1000 |

### 4.1 Cloud Computing Models

Through "School Cloud", ICourse can provide online and offline integrated teaching solutions for colleges and universities. "School Cloud", a cloud online education platform, was launched by NetEase to help schools and enterprises build their online courses, by providing a one-stop solution for the entire process of courses from technical solutions, course content, teaching management, and big data support. Owing to NetEase's 20 years of cloud technology experience, it creates a safe and stable SaaS platform.

XueTang Cloud is an open and reliable education cloud service launched by XueTangX. It provides a one-stop curriculum establishment system with mixed teaching methods, learning tracks, and learning assessment services to meet the needs of learners, teachers, and administrators.

SPOC (Small Private Online Course) is a model that directly chooses online learning resources from the main platform and transforms them into on-campus courses on a small scale [14]. This cloud computing model is widely used among the 22 platforms. For example, CNMOOC offers its SPOC cloud platform to partner universities, with designated domain names and independent homepages. The SPOC can also be used to build new courses from the beginning. Human Health MOOC provides the "SPOC platform + course resource" to colleges and universities, as well as convenient independent teaching management functions such as course content management, self-organized course management, and teaching progress management. During the epidemic, students at vocational colleges can use high-quality resources from ICVE' SPOCs for their independent learning. For teachers, SPOCs from the UOOC Alliance and the UMOOCs platforms provide micro-teaching, live courses, PPT teaching, and audio teaching services free of charge. Specifically, the UMOOCs platform provides teaching materials such as "New Horizon College English (Third Edition)" and "New Standard College English (Second Edition)" for teachers and students. Additionally, GaoXiaoBang offers the SPOC for universities to build credit courses by providing platform construction consultation, operation consultation, feedback reports, etc.

Alibaba Cloud (also known as Aliyun) is adopted by the Chinese Moocs platform. It is a subsidiary of Alibaba Group and responsible for cloud computing technology. Alibaba Cloud provides cloud computing services to online businesses and Alibaba's e-commerce ecosystem.

Huike Cloud is built by Huike Cloud Big Data Comprehensive Laboratory and Huawei Cloud Artificial Intelligence Lab. It is used by GaoXiaoBang platform by offering high-quality computing resources and first-hand industry cases. Huike cloud also employs big data and artificial intelligence applications.

Chinese Accounting Net School uses "ZhengBao Cloud" to provide course resources, distance teaching platforms, and teaching support services to universities and colleges across the country. During the pandemic, ZhengBao Cloud, as a public cloud, has provided technical support, learning data storage and analysis, as well as consultation services.

"Experimental Cloud" is the public cloud for experimental virtual teaching projects launched by iLab-X, with its aims focused on helping universities build a virtual shared application cloud environment. It solves practical problems such as insufficient network resource support, high information security guarantee cost, limited online experimental teaching environment, and difficulty in data collection for shared applications. The "Experiment Cloud" service achieves intensive and transparent management of projects by gathering and integrating software and hardware resources, effectively improving the continuous service capability of the experimental virtual teaching project, and incubating provincial and national virtual simulations for schools.

### 4.2 Cloud Computing Tools

The MOOC, short for Massive Open Online Course, includes educational videos, text content, discussion forums and relevant exercises, with the aim of unlimited participation and free access via the internet [1]. It was used by ICourse in 2019. Instructors can conduct live teaching with the MOOC applet and implement student sign-in and roll call through the WeChat applet. Additionally, ICourse uses other cloud computing tools such as quizzes and discussions to examine students' learning effects and facilitate adjustment of teaching strategies.

Rain Classroom is another cloud computing tool, which is jointly developed by Tsinghua University and XueTangX. It uses IT tools such as PPTs integration and WeChat. Faculty can teach online using their own PPTs or existing PPTs from public resources without changing their teaching routines. Students can participate in online learning through WeChat. By Rain Classroom, instructors can develop courseware with MOOC videos and exercises. Teachers can also interact with students by using questions, barrage, and random roll calls. Based on big data from homework feedback and peer learning, teachers can better understand students' current learning states and students can realize their own learning progress.

Superstar Erya covers different cloud computing tools. For example, Superstar Erya provides synchronized tools such as check-in, questionnaires, and answering questions for teachers and students. LMS tools such as online discussions, online exams, and online interactions are also offered. Students can take learning activities such as watching videos, taking chapter quizzes, and reading related books and materials online. In addition, Superstar Erya provides social networking tools based on big data such as providing timely feedback on the users' learning progress through WeChat.

Human Health MOOC uses the cloud computing tool "Chinese medical education question bank" to enrich its examinations resource. This question bank consists of two parts: the first-class test questions satisfy the requirements of the large-scale examinations for graduation. The second-class test questions meet the daily teaching needs.

The CNMOOC has integrated synchronized tools such as peer-review sessions, LMS tools such as supporting online learning and mobile classroom and online course management, as well as social networking tools such as brainstorming. More specifically, LMS tools include supporting online learning for students (video viewing, courseware browsing, notes, assignments, etc.), as well as teaching material preparation and classroom teaching activities for professors (check-in, classroom tests, rush answers, relevant data about online learning behavior, voting, etc.). In addition, social networking tools can be used for discussion and brainstorming on certain topics.

RongYouXueTang guarantees teachers and students to carry out class construction and online learning at home at any time by using different cloud computing tools. For example, by using LMS tools, teachers can open live channels on the platform and display text, pictures, videos, PPT, blackboard writing, and other teaching resources in real-time. They can also communicate with students through social networking tools for discussion and question answering.

"Vocational Education Cloud" of ICVE platform provides comprehensive cloud computing tools for staff in vocational colleges to achieve online teaching goals. For example, data support is used for teachers to manage the learning tasks of each student in each course. College administrators control the operation of all online courses through the analysis of relevant data. If vocational colleges tend to use MOOCs instead of building their own SPOCs, they can directly guide students to ICVE MOOCs for independent learning. Students can obtain learning certificates after completing ICVE MOOCs, and present the certificates to the college as evidence for further credit identification.

e-HuiLearning provides different cloud computing tools to build high-quality curriculum resources and information application systems. It integrates social networking tools for student communication and LMS tools (e.g., assignments, examinations, answering questions and discussions, earning credits, evaluating interactive teaching activities, etc.). In addition, e-HuiLearning establishes a unified certification system for mutual credits recognition.

iLab-X provides zero-distance contact with artificial intelligence, human-computer interaction, supercomputing, virtual reality, augmented reality, cloud computing and other digital cutting-edge technologies. It also serves users in the form of open sharing and provides review functions such as scoring, collection, likes, and comments.

EduCoder provides cloud computing tools for incorporating an intelligent evaluation and operation mechanism into online education. LMS tools effectively support online teaching and experiment of relevant computer courses. Social networking tools are introduced for large-scale integration of learning, practice, and evaluation. For example, teachers and students can carry out online teaching management, experiments, and practical training, as well as live communication through WeChat applets.

## 5 CONCLUSION

One of the major contributing factors to the Chinese success in containing the pandemic was managing the largest concentration of young people at schools. Thus, it is worth taking a close examination of how China has facilitated continuous education of young people while mitigating the pandemic.

On January 27[th], 2020, at an early stage of the pandemic, the State Council of China announced that the start of the spring semester for all schools, from kindergarten to university, would be postponed until further notice. The Chinese Ministry of Education (MOE) decided to switch in-person classroom education to online education based on 22 online platforms for higher education. For online education of Chinese universities during the pandemic, cloud computing has been the backbone infrastructure for 22 online platforms [2, 18].

Cloud computing provides educational services that can be accessed anytime, anywhere, and for any device [12]. The current COVID-19 pandemic has forced many educational institutions and business organizations to adopt cloud-based solutions. China has been the global leader in applying cloud computing to online education during the chaotic COVID-19 pandemic period.

While cloud computing has been applied to supporting distance education and remote learning, there is a paucity of detailed investigation of the platforms deployed for online university education. In this study, we applied the multiple-case study methodology to examine the details of the 22 online educational platforms that have effectively provided education-related services for universities in China.

While the results of this study provide useful information to many higher education institutions about the use of cloud computing-based platforms during the pandemic period, it has several limitations. First, each country has its unique educational system with different support systems, faculty resources, curricula, and students. The Chinese higher education system, with its large scale of educational ICT infrastructure that serves 30.3 million university students, may not be appropriate to other comparatively smaller countries. Second, the current pandemic period has been relatively short, about 24 months in duration thus far. Thus, the accumulated experience and knowledge about the effectiveness of cloud computing-based platforms for online education may have limited validity. Third, we could add Artificial Intelligence (AI) to improve the performance of cloud computing-based platforms. As the potency of AI-driven cloud computing is more productive, reliable, and strategic, its impact on educational efficiency may have exponential consequences. Such potential of AI-enhanced online educational platforms should be further explored with the accumulated experience and knowledge during the pandemic. However, these limitations provide opportunities for meaningful future research, especially a comparative study of the topic in multiple countries.

**ACKNOWLEDGMENTS**

The research of Dr. Hui Han is funded by the European Research Consortium for Informatics and Mathematics (ERCIM) Alain Bensoussan Fellowship Programme and the Fraunhofer Institute for Experimental Software Engineering IESE.

A   **APPENDICES**

1. Icourse www.icourse163.org
2. XueTangX www.xuetangx.com
3. Wisdom Tree Network www.zhihuishu.com
4. Xueyin Online http://xueyinonline.com/
5. Superstar Erya http://erya.mooc.chaoxing.com/
6. Human Health MOOC www.pmphmooc.com
7. UOOC Alliance www.uooc.net.cn
8. CNMOOC www.cnmooc.org
9. RongYouXueTang www.livedu.com.cn
10. Chinese Moocs www.chinesemooc.org
11. UMOOCs http://moocs.unipus.cn
12. GaoXiaoBang https://imooc.gaoxiaobang.com/
13. Ulearing www.ulearning.cn
14. PeopleMooc http://mooc.people.cn/publicCourse/index.html#/index/portal
15. ICVE www.icve.com.cn
16. Chinese Accounting Net School http://chinaacc.edu.chinaacc.com/
17. ZhengBao Cloud Class https://edu.netinnet.cn/
18. Zhejiang Institutions of Higher Learning Online Open Course Sharing Platform www.zjooc.cn
19. e-HuiLearning www.ehuixue.cn
20. Chongqing Universities Online Open Course Platform www.cqooc.com
21. iLab-X www.ilab-x.com
22. EduCoder www.educoder.net